\documentstyle [12pt,axodraw] {article}

\catcode`@=12
\begin{document}
\thispagestyle{empty}
\begin{flushright} UCRHEP-T260\\July 1999\
\end{flushright}
\vskip 0.5in
\begin{center}
{\Large \bf Nearly Mass-Degenerate Majorana Neutrinos:\\
Double Beta Decay and Neutrino Oscillations\\}
\vskip 1.5in
{\bf Ernest Ma\\}
\vskip 0.1in
{\sl Physics Department, Univeristy of California,\\}
{\sl Riverside, CA 92521, USA\\}
\end{center}
\vskip 2.0in
\begin{abstract}\
Assuming equal tree-level Majorana masses for the standard-model neutrinos, 
either from the canonical seesaw mechanism or from a heavy scalar triplet, 
I discuss how their radiative splitting may be relevant to neutrinoless 
double beta decay and neutrino oscillations.
\end{abstract}
\vskip 0.5in

\noindent $\bullet$ Talk given at the International Conference on 
Non-Accelerator New Physics, Dubna, Russia (June 28 - July 3, 1999).

\newpage
\baselineskip 24pt

\section {Introduction}

In this talk I will first discuss\cite{1} two equally simple mechanisms for 
small Majorana neutrino masses, one famous\cite{2} and one not so 
famous\cite{3}.  I will then mention briefly how they are related to 
neutrinoless double beta decay and neutrino oscillations.  My main focus 
will be on the possibility of nearly mass-degenerate neutrinos and their 
radiative splitting due to the different charged-lepton masses.  In 
particular, I show\cite{4} how a two-fold neutrino mass degeneracy can be 
stable against radiative corrections.  I finish with three examples: (1) 
a two-loop explanation\cite{5} of vacuum $(\Delta m^2)_{sol}$, (2) a one-loop 
connection\cite{6} between $(\Delta m^2)_{atm}$ and vacuum 
$(\Delta m^2)_{sol}$, and (3) a one-loop explanation\cite{7} of small-angle 
matter-enhanced $(\Delta m^2)_{sol}$ with the prediction 0.20 eV $< m_\nu <$ 
0.36 eV.

\section {Origin of Neutrino Masses}

In the standard model, leptons are left-handed doublets $(\nu_i, l_i)_L 
\sim (1,2,-1/2)$ and right-handed singlets $l_{iR} \sim (1,1,-1)$ under 
the gauge group $SU(3)_C \times SU(2)_L \times U(1)_Y$.  The absence of 
the gauge singlet $\nu_{iR} \sim (1,1,0)$ implies that $m_{\nu_i} = 0$. 
However, since the Higgs scalar doublet $\Phi = (\phi^+,\phi^0) \sim 
(1,2,1/2)$ exists, there is a unique 5-dimensional operator\cite{8}
\begin{equation}
\Lambda^{-1} \phi^0 \phi^0 \nu_i \nu_j
\end{equation}
for nonzero Majorana neutrino masses.  The underlying theory which realizes 
this operator is usually assumed to be that of the seesaw mechanism\cite{2}. 
In other words, the gauge-invariant operator
\begin{equation}
(\phi^0 \nu_i - \phi^+ l_i) (\phi^0 \nu_j - \phi^+ l_j)
\end{equation}
is obtained by inserting a heavy Majorana fermion singlet $N$ as the 
intermediate state, as illustrated in Fig.~1 below.

\begin{center}
\begin{picture}(160,80)(0,0)
\ArrowLine(20,40)(80,25)
\Text(50,45)[c]{$\nu_i$}
\DashArrowLine(140,40)(80,25)6
\Text(110,45)[c]{$\phi^0$}
\Line(80,25)(80,-25)
\Text(85,0)[l]{$N$}
\ArrowLine(20,-40)(80,-25)
\Text(50,-20)[c]{$\nu_j$}
\DashArrowLine(140,-40)(80,-25)6
\Text(110,-20)[c]{$\phi^0$}
\end{picture}
\vskip 0.8in
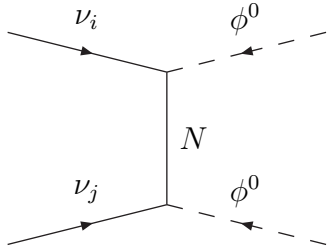
{\bf Fig.~1.} ~ Tree-level realization of the effective operator (2) with 
heavy fermion singlet.
\end{center}

The resulting neutrino mass matrix is then given by
\begin{equation}
({\cal M}_\nu)_{ij} = - {f_i f_j v^2 \over M},
\end{equation}
where $f_i$ are Yukawa couplings of $\nu_i$ to $N$, $v = \langle \phi^0 
\rangle$, and $M$ is the mass of $N$.  On the other hand, the expression 
in (2) can be rewritten as\cite{1,3}
\begin{equation}
\phi^0 \phi^0 \nu_i \nu_j - \phi^+ \phi^0 (\nu_i l_j + l_i \nu_j) + \phi^+ 
\phi^+ l_i l_j,
\end{equation}
which allows the insertion of a scalar triplet $(\xi^{++}, \xi^+, \xi^0)$ as 
the intermediate state, as illustrated in Fig.~2 below.

\begin{center}
\begin{picture}(160,80)(0,0)
\ArrowLine(0,40)(40,0)
\Text(30,30)[r]{$\nu_i$}
\ArrowLine(0,-40)(40,0)
\Text(30,-30)[r]{$\nu_j$}
\DashArrowLine(120,0)(40,0)6
\Text(80,10)[c]{$\xi^0$}
\DashArrowLine(160,40)(120,0)6
\Text(130,30)[l]{$\phi^0$}
\DashArrowLine(160,-40)(120,0)6
\Text(130,-30)[l]{$\phi^0$}
\end{picture}
\vskip 0.8in
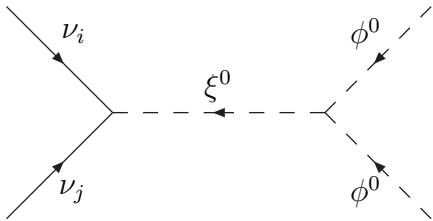
{\bf Fig.~2.} ~ Tree-level realization of the effective operator (4) with 
heavy scalar triplet.
\end{center}

The neutrino mass matrix is now given by
\begin{equation}
({\cal M}_\nu)_{ij} = - {2 f_{ij} \mu v^2 \over M^2},
\end{equation}
where $f_{ij}$ are the Yukawa couplings of $\nu_i$ to $\nu_j$, $\mu$ is the 
trilinear coupling of $\xi$ to $\Phi \Phi$, and $M$ is the mass of $\xi$.  
The alternative way to understand this mass is to note that $\xi^0$ acquires 
a nonzero vacuum expectation value in this model given by $u = -\mu v^2/M^2$. 
In other words, in the limit where $M^2$ is positive and large, it is natural 
for $u$ to be very small.  This method for generating small Majorana neutrino 
masses is as simple and economical as the canonical seesaw mechanism.  To 
obtain the most general $3 \times 3$ neutrino mass matrix, we need 3 $N$'s 
in the latter, but only one $\xi$ in the former. 

\section {Neutrinoless Double Beta Decay and Neutrino Oscillations}

Let the $(\nu_e, \nu_\mu, \nu_\tau)$ mass matrix $\cal M$ have eigenvalues 
$m_{1,2,3}$ with $\nu_e = \sum_i U_{ei} \nu_i$, then
\begin{equation}
{\cal M}_{ee} = \sum_i U_{ei} m_i U^T_{ie} = \sum_i U_{ei}^2 m_i
\end{equation}
is what is being measured in neutrinoless double beta decay.  The most 
recent result from the Heidelberg-Moscow experiment is\cite{9} ${\cal M}_{ee} 
< 0.2$ eV.  Note however that since $U_{ei}^2 m_i$ may be of either sign for 
each $i$, ${\cal M}_{ee}$ does not constrain $|m_i|$ without further 
information.  For example, consider
\begin{equation}
\cal M = \left[ \begin{array} {c@{\quad}c} \cos^2 \theta m_1 + \sin^2 \theta 
m_2 & \sin \theta \cos \theta (m_2 - m_1) \\ \sin \theta \cos \theta (m_2 - 
m_1) & \sin^2 \theta m_1 + \cos^2 \theta m_2 \end{array} \right],
\end{equation}
which tells us that $\nu_e = \cos \theta \nu_1 + \sin \theta \nu_2$.  Now if 
$m_1 > 0$, $m_2 > 0$, then $m_1 < {\cal M}_{ee}$; but if $m_1 < 0$, $m_2 > 
|m_1|$, then there are no individual upper bounds on $|m_1|$ or $m_2$.

In neutrino oscillations, the parameters accessible to experimental 
determination are $\Delta m^2_{ij} = m_i^2 - m_j^2$ and $U_{\alpha i}$, 
hence the sign of $m_i$ is irrelevant there.  The sign of $\Delta m^2_{ij}$ 
is important in matter-enhanced oscillations\cite{10} because neutrino and 
antineutrino forward scattering amplitudes in matter have opposite signs.  

\section {Nearly Mass-Degenerate Majorana Neutrinos and Their Stability 
Against Radiative Corrections}

Suppose neutrinos are Majorana and are equal in mass:
\begin{equation}
\nu_i = U^T_{ie} \nu_e + U^T_{i \mu} \nu_\mu + U^T_{i \tau} \nu_\tau, ~~~ i = 
1,2,3,
\end{equation}
and $m_1 = m_2 = m_3$.  Since $m_e$, $m_\mu$, and $m_\tau$ are all different, 
this degeneracy cannot be exact.  In other words, splitting must occur, but 
how?  This question has two answers.  (1) Depending on the specific mechanism 
by which the neutrinos become massive, there are finite radiative corrections 
to the mass matrix itself\cite{4,5,6}.  (2) There are model-independent 
wavefunction renormalizations which shift the values of the mass matrix 
from one mass scale to another\cite{11}.

The stability of neutrino mass degeneracy against radiative corrections 
depends\cite{4,12} on the symmetry of the mass matrix.  Consider
\begin{equation}
{\cal M} = \left[ \begin{array} {c@{\quad}c} m_{ee} & m_{e \mu} \\ m_{e \mu} 
& m_{\mu \mu} \end{array} \right],
\end{equation}
then
\begin{equation}
\Delta m^2 = (m_{ee} + m_{\mu \mu}) \sqrt {(m_{ee} - m_{\mu \mu})^2 + 
4 m_{e \mu}^2}.
\end{equation}
Thus $\Delta m^2 = 0$ has two solutions.  One is
\begin{equation}
{\cal M} = \left[ \begin{array} {c@{\quad}c} m & 0 \\ 0 & m 
\end{array} \right],
\end{equation}
then the effect of radiative corrections is to shift it by $4 m^2 (\delta_\mu 
- \delta_e)$.  This is inherently unstable.  The other is
\begin{equation}
{\cal M} = \left[ \begin{array} {c@{\quad}c} m & m' \\ m' & -m 
\end{array} \right],
\end{equation}
then the shift is $4 m \sqrt {m^2 + m'^2} (\delta_\mu - \delta_e)$.  This 
is stable as long as $m << m'$ and is easily understood because the 
$m = 0$ limit corresponds to the existence of an extra global $L_e - L_\mu$ 
symmeytry for the entire theory.

\section {Two-Loop Example}

Choose the canonical seesaw mechanism for obtaining neutrino masses.  Impose 
a global $SO(3)$ symmetry so that $(\nu_i,l_i)_L$ and $N_{iR}$ with 
$i = +,0,-$ are triplets. Invariants are then
\begin{equation}
f[(\bar \nu_+ N_+ + \bar \nu_0 N_0 + \bar \nu_- N_-) \bar \phi^0 - (\bar l_+ 
N_+ + \bar \l_0 N_0 + \bar l_- N_-) \phi^-] + h.c.
\end{equation}
and
\begin{equation}
M(2N_+N_- - N_0 N_0).
\end{equation}
Assume $SO(3)$ invariance for $f$ to be valid 
at the electroweak symmetry breaking scale [i.e. no renormalization correction 
from different $\nu_i$'s.]  Let $m_D = f \langle \bar \phi^0 \rangle << M$ 
and $m_0 = m_D^2/M$, then
\begin{equation}
{\cal M}_\nu = \left[ \begin{array} {c@{\quad}c@{\quad}c} 0 & -m_0 & 0 \\ 
-m_0 & 0 & 0 \\ 0 & 0 & m_0 \end{array} \right]
\end{equation}
in the basis $(\nu_+, \nu_-, \nu_0)$.  Now choose $l_+ = e$ so that 
${\cal M}_{ee} = 0$, and let
\begin{equation}
l_- = c \mu + s \tau, ~~~ l_0 = c \tau - s \mu,
\end{equation}
where $c = \cos \theta$, $s = \sin \theta$.

This model\cite{5} differs from the standard model only in the addition of 
3 heavy $N$'s.  The effective low-energy theory differs at tree level only in 
the appearance of 3 nonzero, but equal, neutrino masses.  This degeneracy 
is then lifted in two loops\cite{13}, as illustrated in Fig.~3 below.

\begin{center}
\begin{picture}(360,100)(0,0)
\ArrowLine(0,0)(60,0)
\Text(30,-8)[c]{$\nu$}
\ArrowLine(60,0)(120,0)
\Text(90,-8)[c]{$l$}
\ArrowLine(120,0)(180,0)
\Text(150,-8)[c]{$\nu$}
\Text(180,0)[c]{$\times$}
\ArrowLine(240,0)(180,0)
\Text(180,-12)[c]{$N$}
\Text(210,-8)[c]{$\nu$}
\ArrowLine(300,0)(240,0)
\Text(270,-8)[c]{$l$}
\ArrowLine(360,0)(300,0)
\Text(330,-8)[c]{$\nu$}
\PhotonArc(150,-45)(101,26,154)58
\Text(150,68)[c]{$W$}
\PhotonArc(210,45)(101,206,334)58
\Text(210,-68)[c]{$W$}
\end{picture}
\vskip 1.0in
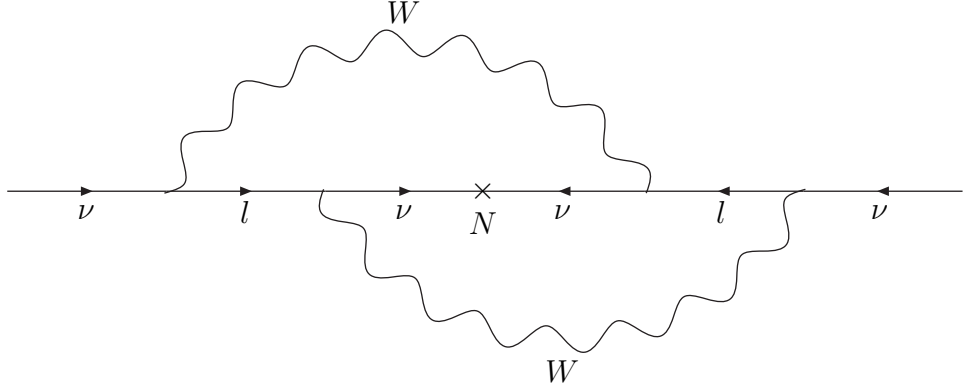
{\bf Fig.~3.} ~ Two-loop radiative breaking of neutrino mass degeneracy.
\end{center}

The leading contribution to the above two-loop diagram is universal, but 
the effects of the charged-lepton masses show up in the propagators, and 
since $m_\tau$ is the largest such mass, the radiative splitting is 
proportional to $m_\tau^2$.  The neutrino mass matrix of Eq.~(15) is now 
corrected to read
\begin{equation}
{\cal M}_\mu = \left[ \begin{array} {c@{\quad}c@{\quad}c} 0 & -m_0-s^2I & 
-scI \\ -m_0-s^2I & 0 & scI \\ -scI & scI & m_0+2c^2I \end{array} \right],
\end{equation}
where
\begin{equation}
I = {g^4 \over 256 \pi^4} {m_\tau^2 \over M_W^2} \left( {\pi^2 \over 6} - 
{1 \over 2} \right) m_0 = 3.6 \times 10^{-9}~m_0,
\end{equation}
and the eigenvalues are $-m_0-s^2I$, $m_0$, and $m_0+(1+c^2)I$.  Let 
$s^2 << 1$, then $\nu_e$ oscillates mostly into $\nu_\mu$ with
\begin{equation}
P(\nu_e \to \nu_e) = {1 \over 2} + {1 \over 2} \cos {\Delta m^2 L \over 2 E},
\end{equation}
where $\Delta m^2 = 2s^2m_0I = 7.2 \times 10^{-9}~s^2 m_0^2 \sim 10^{-10}$ 
eV$^2$, if $s \sim 0.1$ and $m_0 \sim 1$ eV.

This example shows that a minimum splitting of the Majorana neutrino mass 
degeneracy in the canonical seesaw model is suitable for the vacuum 
oscillation solution of the solar neutrino deficit\cite{14}.  However, 
other effects may be larger, such as the renormalization of the $\bar \nu_L 
N_R \bar \phi^0$ vertex.

\section {One-Loop Example I}

Choose the heavy scalar triplet $\xi$ for generating small Majorana neutrino 
masses.  Impose a discrete $S_3$ symmetry, having the irreducible 
representations \underline 2, \underline 1, and \underline 1$'$.  Let 
$(\nu_1, \nu_2) \sim$ \underline 2, and $\nu_3 \sim$ \underline 1, then 
\begin{equation}
{\cal L}_{int} = \xi^0 [f_0(\nu_1 \nu_2 + \nu_2 \nu_1) + f_3 \nu_3 \nu_3] 
+ \mu \bar \xi^0 \phi^0 \phi^0 + ...
\end{equation}
Let $\langle \xi^0 \rangle = u = -\mu \langle \phi^0 \rangle^2 / m_\xi^2$, 
then
\begin{equation}
{\cal M}_\nu = \left[ \begin{array} {c@{\quad}c@{\quad}c} 0 & m_0 & 0 \\ 
m_0 & 0 & 0 \\ 0 & 0 & m_3 \end{array} \right],
\end{equation}
where $m_0 = 2f_0u$ and $m_3 = 2f_3u$.  Now choose $\nu_1 = \nu_e$ so that 
again ${\cal M}_{ee} = 0$, and let $\nu_2 = c\nu_\mu - s\nu_\tau$, $\nu_3 = 
c \nu_\tau + s \nu_\mu$.

This model\cite{6} allows the radiative splitting of the two-fold neutrino 
mass degeneracy to occur in one loop, as illustrated in Fig.~4 below.

\begin{center}
\begin{picture}(360,150)(0,0)
\ArrowLine(30,0)(90,0)
\Text(60,-8)[c]{$\nu_i$}
\ArrowLine(180,0)(90,0)
\Text(135,-8)[c]{$\tau_L$}
\ArrowLine(270,0)(180,0)
\Text(225,-8)[c]{$\tau_R$}
\ArrowLine(330,0)(270,0)
\Text(300,-8)[c]{$\nu_\tau$}
\DashArrowLine(180,-40)(180,0)6
\Text(180,-50)[c]{$\langle \phi^0 \rangle$}
\DashArrowLine(180,97)(180,57)6
\Text(180,106)[c]{$\langle \phi^0 \rangle$}
\DashArrowArcn(180,-45)(101,154,90)6
\Text(130,52)[c]{$\xi^-$}
\DashArrowArcn(180,-45)(101,90,26)6
\Text(240,52)[c]{$\phi^-$}
\end{picture}
\vskip 1.0in
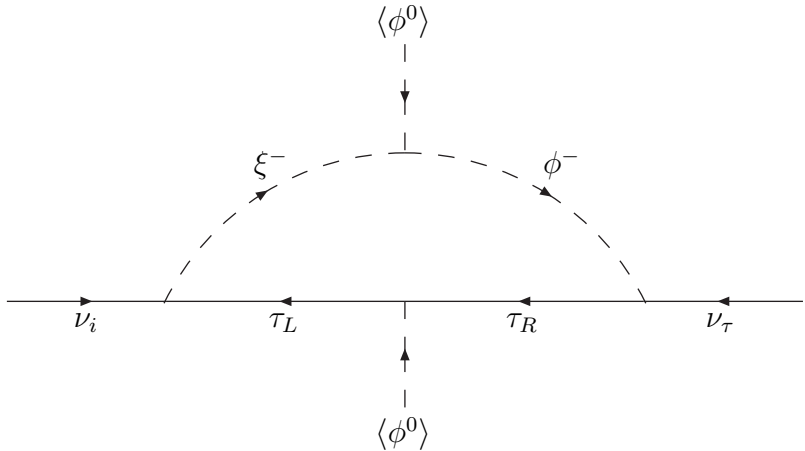
{\bf Fig.~4} ~ One-loop radiative breaking of neutrino mass degeneracy.
\end{center}

Hence ${\cal M}_\nu$ of Eq.~(21) becomes
\begin{equation}
{\cal M}_\nu = \left[ \begin{array} {c@{\quad}c@{\quad}c} 0 & m_0(1+s^2I) & 
-scm_0I \\ m_0(1+s^2I) & 0 & -scm_3I \\ -scm_0I & -scm_3I & m_3(1+2c^2I) 
\end{array} \right],
\end{equation}
whose eigenvalues are $m_3(1+2c^2I)$, and
\begin{equation}
\mp m_0(1+s^2I) \mp {s^2 c^2 (m_0 \mp m_3)^2 I^2 \over 2 (m_0 \pm m_3)},
\end{equation}
with
\begin{equation}
I = \left( {1 \over 4 \pi^2} - {1 \over 16 \pi^2} \right) {G_F m_\tau^2 
\over \sqrt 2} \ln {m_\xi^2 \over M_W^2},
\end{equation}
where the second term inside the parentheses comes from the shift of the 
neutrino wavefunction renormalization from $m_\xi$ to $M_W$.  Numerically, 
$I^2 << (m_o-m_3)^2/(m_0+m_3)^2$, hence
\begin{equation}
\Delta m^2_{12} \simeq {8 s^2 c^2 I^2 m_\nu^4 \over m_0^2 - m_3^2},
\end{equation}
where $m_\nu \simeq m_0 \simeq m_3$ has been used.  Thus a simple 
connection between atmospheric\cite{15} and solar neutrino vacuum 
oscillations is obtained:
\begin{equation}
{(\Delta m^2)_{sol} (\Delta m^2)_{atm} \over m_\nu^4 (\sin^2 2 \theta)_{atm}} 
= 2I^2 = 4.9 \times 10^{-13} \left( \ln {m_\xi^2 \over M_W^2} \right)^2.
\end{equation}
This equality holds for the sample values of $m_\nu = 0.6$ eV, $(\sin^2 2 
\theta)_{atm} = 1$, $m_\xi = 1$ TeV, $(\Delta m^2)_{sol} = 4 \times 10^{-10}$ 
eV$^2$, and $(\Delta m^2)_{atm} = 4 \times 10^{-3}$ eV$^2$.  [If $m_\xi = 
10^{13}$ GeV, then $m_\nu \sim 0.2$ eV.]

This example shows that it is possible to have a one-loop effect, but which 
appears only in second order because of nondegenerate $(m_0 \neq m_3)$ 
perturbation theory.  In the previous example, the effect is two-loop but it 
occurs in first order because of degenerate perturbation theory.

\section {One-Loop Example II}

This model\cite{7} is a variation of Example I, with $\nu_1 \nu_1 + \nu_2 
\nu_2$ as an invariant, say under $SO(2)$.  Hence
\begin{equation}
{\cal M}_\nu = \left[ \begin{array} {c@{\quad}c@{\quad}c} m_0 & 0 & 0 \\ 
0 & m_0(1+2c^2I) & -sc(m_0+m_3)I \\ 0 & -sc(m_0+m_3)I & m_3(1+2s^2I) 
\end{array} \right],
\end{equation}
where $\nu_1 = \nu_e$, $\nu_2 = c \nu_\tau - s \nu_\mu$, $\nu_3 = c \nu_\mu + 
s \nu_\tau$.  Now rotate $\nu_1$ and $\nu_2$ slightly by $\theta'$, then the 
small-angle matter-enhanced solution to the solar neutrino deficit works for 
$\sin^2 2 \theta' \simeq (2 - 10) \times 10^{-3}$ and
\begin{equation}
(\Delta m^2)_{12} = 4 c^2 I m_0^2 \simeq (3 - 10) \times 10^{-6}~{\rm eV}^2.
\end{equation}
For $c^2 = 0.7$, i.e. $(\sin^2 2 \theta)_{atm} = 0.84$, and $m_\xi = 10^{14}$ 
GeV, this implies
\begin{equation}
0.20 ~{\rm eV} < {\cal M}_{ee} < 0.36 ~{\rm eV}.
\end{equation}
Experimentally, the most recent Heidelberg-Moscow result\cite{9} is 
${\cal M}_{ee} < 0.2$ eV, but the expected sensitivity is only 0.38 eV, 
both at 90\% confidence level.  More data may see something or rule out the 
above prediction.

\section {Conclusions}

\noindent $\bullet$ Neutrino mass is equally natural coming from the seesaw 
mechanism or a heavy scalar triplet.

\noindent $\bullet$ If $\nu_{e, \mu, \tau}$ are nearly mass-degenerate, their 
radiative splitting may be suitable for solar neutrino oscillations.  Details 
depend on the specific model, but the smallness of vacuum $(\Delta m^2)_{sol}$ 
is only obtained in certain special cases.

\section*{Acknowledgements}

I thank Sergey Kovalenko and everyone associated with the organization of 
NANP-99 for their great hospitality.  This work was supported in part by the 
U.~S.~Department of Energy under Grant No.~DE-FG03-94ER40837.

\bibliographystyle{unsrt}

\end{document}